# A Digital Calibration Source for 21 cm Cosmology Telescopes

Kalyani Bhopi[1,2,5], Will Tyndall[3], Pranav Sanghavi[1,2], Kevin Bandura[1,2],
Laura Newburgh[3] and Jason Gallicchio[4]

[1]Lane Department of Computer Science and
Electrical Engineering, Morgantown, WV, USA

[2]Center for Gravitational Waves and Cosmology
West Virginia University, Morgantown, WV, USA

[3]Department of Physics, Yale, New Haven, CT, USA

[4]Harvey Mudd College, 301 Platt Blvd.
Claremont, CA 91711, USA

[5]kbb00008@mix.wvu.edu

Foreground mitigation is critical to all next-generation radio interferometers that target cosmology using the redshifted neutral hydrogen 21 cm emission line. Attempts to remove this foreground emission have led to new analysis techniques as well as new developments in hardware specifically dedicated to instrument beam and gain calibration, including stabilized signal injection into the interferometric array and drone-based platforms for beam mapping. The radio calibration sources currently used in the literature are broadband incoherent sources that can only be detected as excess power and with no direct sensitivity to phase information. In this paper, we describe a digital radio source which uses Global Positioning Satellite (GPS) derived time stamps to form a deterministic signal that can be broadcast from an aerial platform. A copy of this source can be deployed locally at the instrument correlator such that the received signal from the aerial platform can be correlated with the local copy, and the resulting correlation can be measured in both amplitude and phase for each interferometric element. We define the requirements for such a source, describe an initial implementation and verification of this source using commercial Software Defined Radio boards, and present beam map slices from antenna range measurements using the commercial boards. We found that the commercial board did not meet all requirements, so we also suggest future directions using a more sophisticated chipset.

*Keywords*: Radio astronomy; 21 cm cosmology; noise calibration; astronomical instrumentation; beam characterization.

## 1. Introduction

Radio surveys of the redshifted 21 cm emission line of neutral hydrogen promise to measure statistical cosmological signals that cannot be accessed through other means: measurements of the Dark Ages and Cosmic Dawn at redshifts $z \sim 15 - 35$ (e.g. LEDA Price *et al.* (2018)); the Epoch of Reionization at redshifts $z \sim 6 - 15$ (e.g. upcoming or deployed interferometers HERA (Aguirre *et al.*, 2022; DeBoer *et al.*, 2017), LOFAR (Zaroubi & Silk, 2005), SKA (Square Kilometer Array, 2009), MWA (Tingay *et al.*, 2013), PAPER (Backer *et al.*, 2007) and global experiments like EDGES (Monsalve *et al.*, 2017), SARAS3 (Nambissan *et al.*, 2021), PRIZM (Philip *et al.*, 2019)); and measurements of Dark Energy at redshifts $z \sim 1 - 2$ (e.g. from GBT (Chang *et al.*, 2008), CHIME (The CHIME Collaboration, 2022a; Bandura *et al.*, 2014), HIRAX (Crichton *et al.*, 2022; Newburgh *et al.*, 2016)).

[5]Corresponding author.

Some experiments have detected large scale structure in combination with optical surveys (The CHIME Collaboration, 2022b; Wolz *et al.*, 2022; Li *et al.*, 2021; Tramonte & Ma, 2020; Chang *et al.*, 2010; Masui *et al.*, 2010, 2013), placed limits on $\Omega_{HI}$ (Switzer *et al.*, 2013), placed limits on IGM heating at high redshift (Pober *et al.*, 2015) and the 21 cm power spectrum (Kolopanis *et al.*, 2019; Li *et al.*, 2019; Beardsley *et al.*, 2016; Ewall-Wice *et al.*, 2016), and a tension has appeared between results from different global experiments (Bowman *et al.*, 2018; Singh *et al.*, 2022). Simulations and analyses from these experiments have found that the primary challenge for 21 cm measurements is the presence of bright synchrotron foreground emission from the galaxy (e.g. Kerrigan *et al.* (2018), Kohn *et al.* (2016), and Thyagarajan *et al.* (2015)). Suppressing this foreground emission may be possible because the foregrounds are smooth in frequency while the cosmological signal is not. This provides a pathway for foreground mitigation through a variety of possible filters (e.g. Ewall-Wice *et al.* (2021), Morales *et al.* (2019), Shaw *et al.* (2014), Shaw *et al.* (2015), Liu *et al.* (2014), and Parsons *et al.* (2012)), and it has been shown that beam measurements are essential for adequate filtering, particularly for lower redshift surveys (Seo & Hirata, 2016).

Successfully separating signal and foreground components requires knowledge of any frequency dependence in the instrument that could introduce spectral features in the otherwise smooth foreground. The frequency dependence can be accounted for in the cosmological analysis as long as it is well-measured, here we target a 1% requirement for each measurement at a 170 ms integration time, which can be scaled appropriately to meet individual experimental requirements (e.g. Shaw *et al.* (2015)). Because the 21 cm signal is small (∼100 mK) and the instruments are designed not to resolve sources, dedicated intensity mapping instruments are typically compact (the dish–dish spacing is nearly commensurate with the dish diameter), many-element transit interferometers to reduce statistical noise on relevant large scales (Parsons *et al.*, 2019), such that it is difficult to achieve beam and gain measurements using techniques developed for dispersed and steerable interferometers. As a result, many 21 cm experiments are currently developing instrumentation for meeting their calibration goals, for example by flying radio sources on quadcopter drones to measure beam patterns (Zhang *et al.*, 2021; Chang *et al.*, 2015; Virone *et al.*, 2014; Jacobs *et al.*, 2016; Pupillo *et al.*, 2015) or by injecting a calibration signal that is common to multiple antennas in the interferometric array to assess instrument stability (Newburgh *et al.*, 2014).

However, the calibration sources currently used are incoherent (total power only, e.g. (Patra *et al.*, 2017)), and so these instruments must contend with signal-to-noise issues where the beam has low response as well as the lack of a direct phase measurement without a reference antenna (Makhija *et al.*, 2021; Fritzel *et al.*, 2016; Ciorba *et al.*, 2019). In this paper, we describe a new type of radio source in development for radio interferometer calibration, which has the following characteristics: (i) it forms a deterministic calibration signal based on a pseudo-random sequence generated from a time stamp. Such a deterministic signal can be copied and correlated in the radio instrument correlator, allowing for a coherent measurement. Because the free-space calibration source is deterministic, it is suitable for deployment on an aerial platform above the instrument, for example on a cube-sat or quadcopter drone. (ii) The correlated signal can be used to measure both total power (as would be measured from the incoherent sources currently used by 21 cm experiments), and also be used to measure the instrument phase relative to a fixed source (not directly measureable with an incoherent source). This provides an independent cross-check of the telescope-pair (visibility) phases, and also provides a useful data set for model-building and comparing to simulations in the sidelobes where the phase may vary rapidly when measured against a fixed source. Correlation also improves the signal-to-noise ratio (SNR) of the calibration measurement in most measurement regimes, and may be more robust to human-generated radio frequency interference (RFI). (iii) It can have a wide bandwidth and flexible band selection such that it can be adapted for a wide range of 21 cm intensity mapping radio telescopes.

In Sec. 2, we describe the theoretical underpinning and requirements of a digital noise source and in Sec. 3 we describe preliminary results from a commercially available software-defined radio (SDR) board. We conclude in Sec. 4 which includes future directions with more capability than the SDR source used for the measurements presented here.

## 2. Digital Noise Source Overview

### 2.1. *Design concept*

The deterministic calibration source scheme is shown in Fig. 1. The calibration source has two identical copies: one that is transmitted into the telescope (for example from a drone), the other copy is directly attached as an analog input to one of the channels of the radio instrument correlator. The Field-Programmable Gate Array (FPGA) on each calibration source takes in a GPS time stamp once per second and generates a deterministic seed from that time stamp. That seed is used to produce a sequence of bits which is random, but entirely defined by the time-seed value (known as a pseudo-random sequence). The resulting complex (real and imaginary) digital signal is converted to an analog signal by an on-board DAC. Identical signals would be transmitted into the telescope aperture and also generated locally at the correlator because each board is forming the same sequence from the same seed, synchronized by GPS timestamp. The two identical signals (one received through the telescope, and one locally generated) can be correlated in the radio telescope data acquisition system to produce a measured response to the calibration source for each radio array antenna at the same frequency resolution as the full instrument. The correlated calibration signal is complex (amplitude and phase) and also deterministic because the signal itself is determined by the GPS time stamps.

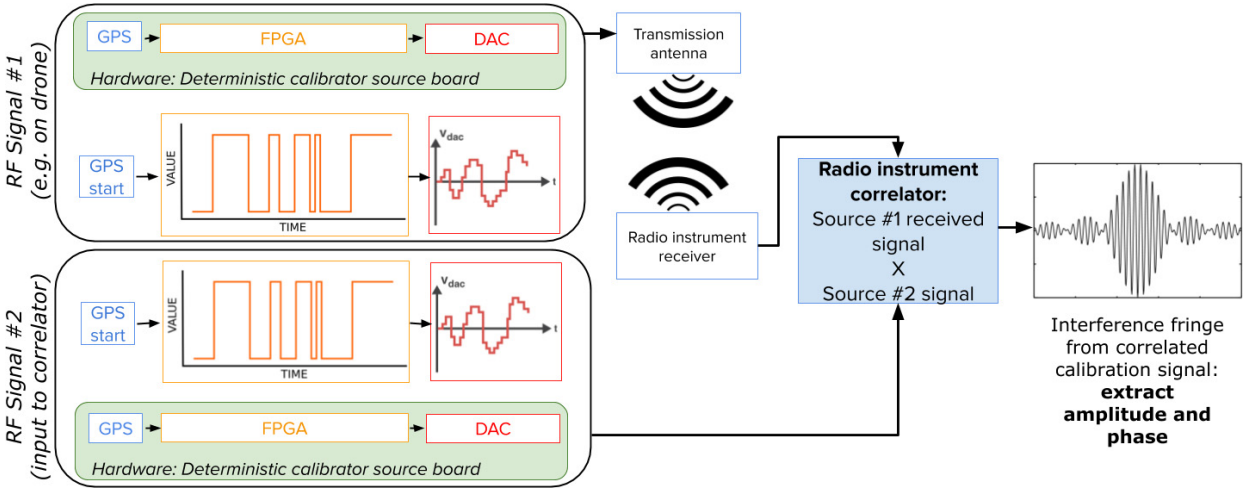

Fig. 1. (Color online) Illustration of the digital calibration source concept, described in more detail the text. This shows a generic signal generated in the FPGA from a time stamp and converted in the DAC. In practice the choice of pseudo-random sequence and relative real and imaginary components of the signal will be optimized for robustness to systematics.

In the literature, pseudo-random noise signal generation for calibration has been proposed and measured once (Perez *et al.*, 2009), specifically for oceanography-based applications at high frequencies (1.57 GHz) and low bandwidth (2 MHz) compared to cosmological 21 cm telescopes. The authors tested the signal generation with a custom board and compared the correlated results to simply splitting a signal (which should have perfect correlation). They showed that the two were identical, validating the approach of using pseudo-random numbers as a calibration standard. The authors noted that pseudo-random noise has the benefit of acting as thermal noise and hence can be used as a broad-band calibrator and suggested this signal could be injected in radio interferometers through a large coaxial cable signal distribution network. Digital noise sources have also been used for testing radio telescope backends (Buch *et al.*, 2014), with variable correlation between output signals. The source described in this paper furthers these works by enabling a wider bandwidth source and distributing the signal entirely without cables by developing the critical step of triggering the sequence with a GPS time-stamp to form a fully deterministic signal to enable beam and analog system characterization. The primary new measurement this enables is a measurement of the amplitude and phase of the radio telescope beam using a digital calibration source flown on an aerial platform.

### 2.2. *Digital noise source figure of merit*

Incoherent signals can only be detected as total power, and thus receiver noise creates a fundamental integration time requirement for incoherent measurements. A deterministic source can be measured in both auto- and cross-correlation, which allows both improved signal to noise in many relevant noise regimes as well as a measurement of phase. In this section, we will define a figure of merit based on recovering an input gain term ($g_T$), which also serves as our SNR variable. In later sections, we will simulate this recovery in the presence of noise and timing jitter to identify requirements for SNR and timing jitter. To define the figure of merit, we begin with the signal received through the telescope. We flash the source on and off such that the signals with the calibration signal on and off are given by

$$V_{T(ON)} = g_T(s_{cal} + s_{sky}) + s_{T,noise\ (ON)}, \quad (1)$$

$$V_{T(OFF)} = g_T s_{sky} + s_{T,noise\ (OFF)}. \quad (2)$$

The signal received from the reference calibration source that is directly connected to the correlator is given by

$$V_{ref(ON)} = g_{ref} s_{cal} + s_{ref,noise\ (ON)}, \quad (3)$$

$$V_{ref(OFF)} = s_{ref,noise\ (OFF)}, \quad (4)$$

where $V_T$ is the voltage measured through the telescope receiver, $V_{ref}$ is the voltage measured from a the calibration reference source that is directly connected to the correlator, $s_{cal}$ and $s_{sky}$ are the calibration source signal and sky signal, respectively, and $s_{noise}$ represents the noise added by the system to the received signal. $g_T$ is the overall gain seen by the telescope receiver, here we will use it in two ways: it is a proxy for the SNR and thus used to define the relative values of the signal and noise; it is also the input number we will recover as part of our figure of merit because the beam acts as a time-dependent gain during the course of a drone beam measurement. In this convention, noise is separate from the gain.

The resulting auto-correlation measured through the telescope, and at the correlator from the calibration reference source, is

$$\langle VV^*\rangle_{T(ON)} = \langle g_T g_T^*\rangle(S_{cal} + S_{sky}) + S_{T,noise\ (ON)} + \epsilon_{corr,1}, \quad (5)$$

$$\langle VV^*\rangle_{T(OFF)} = \langle g_T g_T^*\rangle S_{sky} + S_{T,noise\ (OFF)} + \epsilon_{corr,2}, \quad (6)$$

$$\langle VV^*\rangle_{ref(ON)} = \langle g_{ref} g_{ref}^*\rangle S_{cal} + S_{ref,noise\ (ON)} + \epsilon_{corr,3}, \quad (7)$$

$$\langle VV^*\rangle_{ref(OFF)} = S_{ref,noise\ (OFF)}. \quad (8)$$

The resulting cross-correlation between the signal measured through the telescope and the signal from the reference calibration source is

$$\langle V_T V_{ref}^*\rangle_{(ON)} = \langle g_T g_{ref}^*\rangle S_{cal} + S_{corr} + \epsilon_{corr,4}. \quad (9)$$

The time-averaged auto-correlation calibration signal detected through the telescope, $\langle VV^*\rangle_{T(ON)}$ contains the calibration source power ($S_{cal} = \langle s_{cal} s_{cal}^*\rangle$), the sky signal ($S_{sky} = \langle s_{sky} s_{sky}^*\rangle$), the noise signal ($S_{T,noise} = \langle s_{T,noise} s_{T,noise}^*\rangle$), and the term $\epsilon_{corr,1}$ which contains $(g_T^2\langle s_{cal} s_{sky}\rangle)$, $(g_T\langle s_{cal} s_{T,noise(ON)}\rangle)$, $(g_T\langle s_{sky} s_{T,noise(ON)}\rangle)$ cross terms. The time-averaged auto-correlation signal detected through the telescope when calibration source is off, $\langle VV^*\rangle_{T(OFF)}$ contains the sky signal $S_{sky}$, the noise signal $S_{T,noise}$, and the term $\epsilon_{corr,2}$ which contains $(g_T\langle s_{sky} s_{T,noise(OFF)}\rangle)$ cross terms. The time-averaged auto-correlation signal detected from the reference calibration source, $\langle VV^*\rangle_{ref(ON)}$ contains the calibration source power $S_{cal}$, the noise signal ($S_{ref,noise} = \langle s_{ref,noise} s_{ref,noise}^*\rangle$), and the term $\epsilon_{corr,3}$ which contains $(g_{ref}\langle s_{cal} s_{ref,noise(ON)}\rangle)$ cross terms. The cross-correlation between the measured voltage from calibration source and reference source, $\langle V_T V_{ref}^*\rangle_{(ON)}$, contains the calibration signal $S_{cal}$, any correlated noise ($S_{corr}$) from the instrument, which we assume is designed to be negligible, and the term $\epsilon_{corr,4}$ which contains $(g_T g_{ref}\langle s_{sky} s_{cal}\rangle)$, $(g_T\langle s_{cal} s_{ref,noise(ON)}\rangle)$, $(g_T\langle s_{sky} s_{ref,noise(ON)}\rangle)$ cross terms. All the $\epsilon_{corr}$ terms scale as $1/N$, where $N$ is the number of samples taken into consideration, resulting from finite time and bandwidth. Only the regimes of bandwidth and integration time, where these terms can be neglected, are considered.

The auto-correlation and cross-correlation measurements contain noise terms generated by the system noise temperature and gain fluctuations. For the simulations that follow, the variable $g_T$ defines the SNR: $g_T \sim \frac{s_{cal}}{s_{noise}}$ such that the contribution to the measured signal from the noise is small when the gain is large. The specific implementation of this is described in Sec. 2.3. This calibration source is designed to recover the angle-dependent telescope gain (the beam). For a drone-calibration beam mapping campaign, this would be a time-dependent SNR such that $g_{beam} \propto g_T$. Thus, we define a figure of merit based on how well we recover the input $g_T$ term. We use combinations of source on and off data to define the estimator $\hat{g}$ for both the auto- and cross-correlation data sets to recover $g_T$:

For Auto-correlation:

$$\hat{g}_{auto} = \frac{\langle VV^*\rangle_{T(ON)} - \langle VV^*\rangle_{T(OFF)}}{\langle VV^*\rangle_{ref(ON)} - \langle VV^*\rangle_{ref(OFF)}}$$

$$= \frac{\langle g_T g_T^*\rangle}{\langle g_{ref} g_{ref}^*\rangle} + \epsilon_{corr,5}, \quad (10)$$

where the term $\epsilon_{corr,5}$ equals $\frac{\epsilon_{corr,1} - \epsilon_{corr,2}}{\langle g_{ref} g_{ref}^*\rangle S_{cal} + \epsilon_{corr,3}}$ which can be considered an additional noise due to cross terms and gain fluctuations captured in the simulations presented below.

For Cross-correlation:

$$\hat{g}_{\text{cross}} = \langle V_{\text{T}} V_{\text{ref}}^* \rangle_{(\text{ON})}$$

$$= \langle g_{\text{T}} g_{\text{ref}}^* \rangle S_{\text{cal}} + S_{\text{corr}} + \epsilon_{\text{corr},4}. \quad (11)$$

For the simulation realizations presented below, $g_{\text{ref}}$ is chosen to be 1.0 and $S_{\text{cal}}$ has unit standard deviation such that $\hat{g}_{\text{auto}} \simeq \langle g_{\text{T}} g_{\text{T}}^* \rangle$ and $\hat{g}_{\text{cross}} \simeq g_{\text{T}}$. Again, the $\simeq$ is shown because the simulations will include noise, digitization effects, and relative jitter. The choice to set $g_{\text{ref}} = 1.0$ reflects the fact that while the calibration source may vary, we can both maximize its stability and directly measure (and potentially use feedback) to keep the time-dependence significantly lower than that which we are measuring. As a result, we normalize the estimator $\hat{g}$ of the system as follows:

For auto-correlation:

$$\hat{g}_{\text{auto,norm}} = \frac{\hat{g}_{\text{auto}}}{g_{\text{T}}^2}. \quad (12)$$

For cross-correlation:

$$\hat{g}_{\text{cross,norm}} = \frac{\hat{g}_{\text{cross}}}{g_{\text{T}}}, \quad (13)$$

where $g_{\text{T}}$ is the SNR we have chosen for a given simulation realization. In the noiseless limit, these normalized estimators would be precisely 1.0 in the simulations that follow. As a result, these estimators are used as our figure of merit for the recovery of the input $g_{\text{T}}$ which we require to 1% (or better) in power.

### 2.3. *Specifications on noise and timing from figure-of-merit simulations*

We investigate the precision requirements necessary for our desired beam amplitude and phase recovery by simulating the correlation measurements that occur within the IceBoard FPGA deployed on 21 cm cosmology telescopes like CHIME and HIRAX (Bandura *et al.*, 2016). During the simulations, our parameter value choices are motivated by the technical specifications of the IceBoard system, but the results will be applicable for any digital correlator with similar frequency bandwidths ($\mathcal{O}(500\,\text{kHz})$) and integration times ($\mathcal{O}(40\,\text{ms})$). The python code to generate the simulations and figures is publicly available.[a]

---

[a]https://github.com/WTyndall/DNS_Sims.

We simulate the complex time series voltage arrays generated by two different digital noise sources (Eqs. (1)–(4)) and record their lag auto-correlation and cross-correlation (Eqs. (5)–(9)) in a single 390 kHz bandwidth frequency channel. We generate a signal array $s_{\text{cal}}$, and two noise arrays $s_{\text{T,noise(ON)}}$ and $s_{\text{T,noise(OFF)}}$, each separately containing 65,536 ($2^{14}$) complex float values pseudo-randomly drawn from a normal distribution centered around zero with a standard deviation of the real and imaginary components separately equal to 1. This is equivalent to a 169.972 ms (4 integration period) time series. As described, this choice sets $g_{\text{ref}} = 1.0$ and $s_{\text{cal}}$ to have unit standard deviation. Then, $s_{\text{cal}}$ is multiplied by the signal-to-noise parameter $g_{\text{T}}$ that we are free to choose, or equivalently the noise arrays are divided by $g_{\text{T}}$. The noise arrays contain random noise and also absorb the sky signal ($g_{\text{T}} s_{\text{sky}}$).

The voltage arrays are then optionally "quantized" to imitate the digitization process such that they are rounded and rescaled by $128/(6\,\Re(V_{\text{RMS}}))$ to 8-bit complex integer arrays. This scaling parameter is optimized for each choice of the SNR $g_{\text{T}}$ to utilize as much of the 8-bit range as possible when digitized, which forces the $6\sigma_{\text{RMS}}$ to be within the 8-bit range. This choice reduces the impact of quantization errors such that they made no difference to the results, and so are not included in the following analysis. Finally, the auto and cross-correlations are calculated for the quantized and unquantized voltage arrays by multiplying the time streams in Fourier space, resulting in a lag spectrum. For each iteration, we record the maximum complex amplitude of the auto-correlation and cross-correlation (which also includes phase), providing the values defined by Eqs. (5)–(9). These maxima occur in the zero index of the lag spectrum, and the values are equivalent to the correlated power received in a single 390 kHz frequency channel for the full integration period. The maximum amplitude and phase of these correlations are used to calculate the figure of merit defined in Eqs. (12) and (13).

#### 2.3.1. *Noise and quantization*

First, we assess the impact of noise on our ability to recover the beam amplitude, $g_{\text{T}}$. Keeping integration time fixed to the values in the previous section, we vary the gain from 0.005 to 3.2 in increments of 0.05. The low end of this range was chosen to reflect the likely SNRs for sidelobe measurements that also target measurements of the main beam without saturation, which are typically SNR < 1 per integration period. For each chosen $g_T$, we generate 10,000 realizations of the $V_{T(ON)}$, $V_{T(OFF)}$, $V_{ref(ON)}$, and $V_{ref(OFF)}$ signals. For each realization, we compute the normalized estimator for both the auto and cross-correlations (Eqs. (12) and (13)). The mean is computed across all realizations and stored as the estimator $\hat{g}_{norm}$ for both auto- and cross-correlations. The standard deviation is also computed and divided by the mean to provide the error on the estimator. The results are shown in Fig. 2. Figure 2(left) shows the mean values of the estimator, which indicates that for low values of $g_T$ (low SNR) the auto-correlation has significantly higher fluctuations than the cross-correlations, and they begin to converge around SNRs of 3. Figure 2(right) shows the standard deviations across the realizations, divided by the mean, providing a percentage error on the beam recovery in a smaller range of SNR ($g_T < 2$). That figure also includes the 1% error goal and indicates the SNR at which the cross-correlation and auto-correlations each meet the 1% goal.

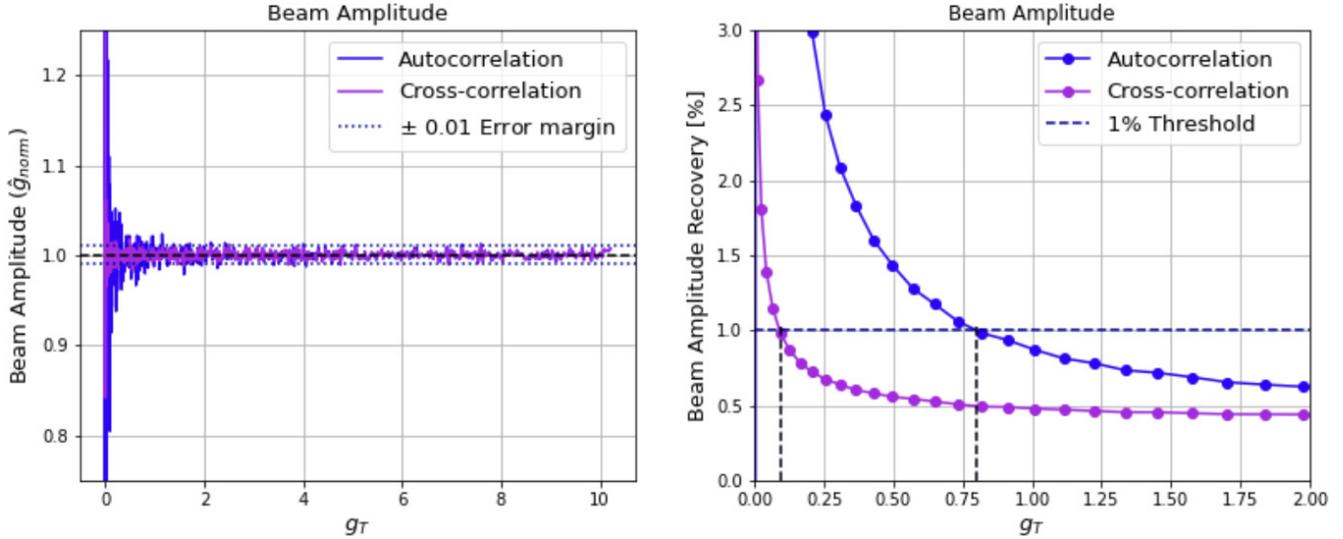

Fig. 2. (Color online) Simulation results showing recovered beam amplitude (left) and beam amplitude recovery error (right) as a function of SNR for auto-correlation and cross-correlation simulated measurements. A horizontal line has been drawn at 1% to show where the beam recovery meets our requirement. The beam amplitude recovery error curve goes below the 1% requirement at $g_T = 0.79$ for auto-correlation and $g_T = 0.09$ for cross-correlation.

Table 1. Figure-of-merit for the telescope receiver system.

| Parameter | Requirement | Specification |
| --- | --- | --- |
| SNR | auto-correlation beam amplitude recovery to 1% | $g_T \geq 0.79$ |
| SNR | cross-correlation beam amplitude recovery to 1% | $g_T \geq 0.09$ |
| Timing jitter | cross-correlation beam amplitude recovery to 1% at $g_T = 0.73$ | $\Delta t < 1.7$ ns |
| Timing jitter | cross-correlation beam amplitude recovery to 1% at $g_T = 0.12$ | $\Delta t < 0.7$ ns |
| Timing jitter | phase recovery to 1% at $g_T = 0.12$ | $\Delta t < 2.7$ ns |

From the simulations varying the SNR ratio via varying $g_T$, the beam recovery is better for the cross-correlation measurement than the auto-correlation measurement in all regimes, but they approach each other at high values of $g_T$, as expected. In addition, the cross-correlation meets the 1% requirement at any input gain values above 0.09, which indicates that 1% measurements are feasible even for low-signal regimes such as sidelobe regions in the beam. The autocorrelation measurement can also achieve 1% measurements at gain values of 0.79. These results are summarized in Table 1.

2.3.2. *Timing*

The telescope calibration procedure outlined in this paper will require two unique clock signals to seed the distributed digital noise sources. This results in static timing offsets ($\mathcal{O}(1\ \mu s)$ due to the drone flight distance and cable delays) and variable timing jitters (between 1 ps and 4 ns) between the two clock signals. To provide a specification for the precision required between the two clocks, we compute the $\hat{g}_{\text{norm}}$ while varying the timing jitter and SNR $g_T$ to investigate different regimes.

To estimate the effect of timing jitter from the calibration source, we will assume the clock seeding the noise source connected directly to the correlator $V_{\text{ref(ON)}}$ has negligible intrinsic jitter, and thus the phase and index shifts occur only for the signal observed by the telescope. We model the difference in clock time for the digital noise source onboard the drone $V_{\text{T(ON)}}$ for each value of $i$ in the 65,536 element time series using:

$$\Delta t_i = t_{\text{T},i} - t_{\text{ref},i} = \Delta t_{\text{FE}} + \Delta t_{\text{FS}} + \delta t_i \simeq \delta t_i, \quad (14)$$

where $\Delta t_{\text{FE}}$ is the front-end time delay from signals traveling through the coaxial cables, $\Delta t_{\text{FS}}$ is the free-space time delay associated with increased path length due to the geometry of the drone, and $\delta t_i$ is the random jitter with a varying magnitude. We estimate that the contributions from the front-end and free-space delays are not small, typically $\mathcal{O}(1\ \mu s)$, but we ignore these terms in the following simulations because they can be measured and accounted for. The remaining timing differences are due to the clock jitter, which will result in a phase shift, $\phi$, that depends on the central frequency $f_{\text{cent}}$ of the chosen frequency bin, as

$$\phi_i = f_{\text{cent}} \Delta t_i. \quad (15)$$

The effects of this phase shift are introduced by element-wise multiplication of the complex phase $e^{-j\phi_i}$ and the simulated voltage array $V_{\text{T(ON)},i}$.

To assess the impact of jitter on the correlation measurements, we vary the standard deviation of the Gaussian distribution from which the jitter values are drawn from 1 ps to 4 ns (Mena-Parra *et al.*, 2022). Additionally, we vary the input signal-to-noise parameter $g_T$ from 0.05 to 4.0 in voltage (0.0025 to 16.0 in power). For each value of jitter and gain, 10,000 iterations of auto- and cross-correlations are simulated from unique signal and noise arrays.

The results of these simulations are shown in Fig. 3. For the two gain regimes described in Sec. 2.3.1, we find that the cross-correlation measurements recover the beam amplitude with more precision than the corresponding auto-correlation measurements at low jitter. These two regimes are shown in black in the corresponding figure. The first regime targets a 1% error in beam amplitude recovery for the cross-correlation, which is achieved for an input SNR $g_T = 0.12$ when the jitter is below 0.7 ns. The auto-correlation in this signal-to-noise regime recovers the beam with an error exceeding 5%. The second regime we consider targets a 1% error in beam amplitude recovery for the auto-correlation, which is achieved when $g_T = 0.79$, as shown in Sec. 2.3.1. The closest signal-to-noise value used in the timing simulations is $g_T = 0.73$, for which the cross-correlation achieves 1% precision when the jitter is below 1.7 ns. If $g_T = 0.73$ is held constant while the jitter improves to below 100 ps, the cross-correlation would recover the beam amplitude to around 0.5%. In the entire parameter space explored by the simulations, phase is always recovered to better than 1% accuracy in the cross-correlation for jitter values that satisfy the 1% beam amplitude precision requirement.

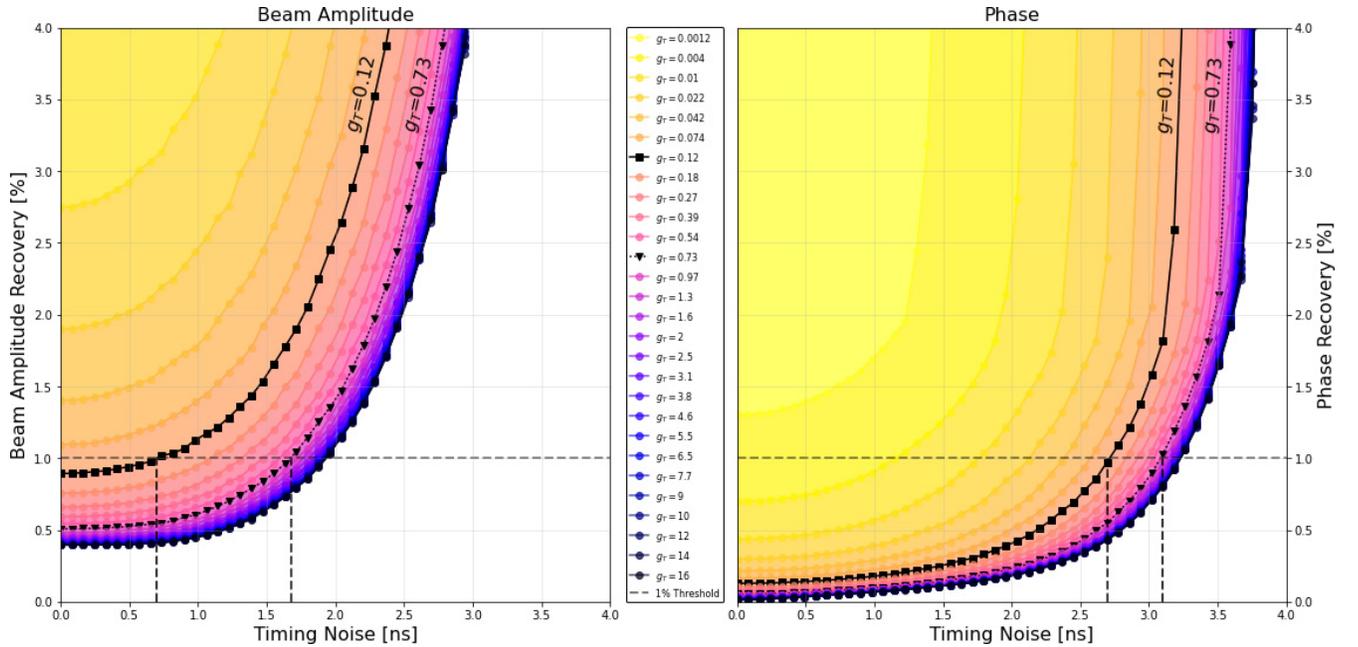

Fig. 3. (Color online) Beam amplitude (left) and phase (right) recovery percent error as a function of the timing noise (jitter, $\delta t$) between correlated signals in digital noise source simulations. Each colored curve contains points obtained from cross-correlation measurements with fixed signal-to-noise parameter ($g_T$) as the timing noise is increased. Two curves ($g_T = 0.12$ and $g_T = 0.73$) are shown in black for enhanced contrast because these values are closest to the gain regimes where the cross-correlation and auto-correlation (respectively) reach 1% beam amplitude recovery precision. A 1% error threshold (and its intersection with the enhanced contrast curves) is shown as a dashed black line.

The combined requirements from Sec. 2.3.1 and jitter simulations are summarized in Table 1. Section 2.3.1 found that the minimum SNR to achieve a 1% measurement in cross-correlation for a ∼170 ms integration period is $g_T = 0.09$, which was computed for zero jitter. Adding jitter changes the minimum SNR to achieve this goal in the cross-correlation. Jitter values of 1.7 ns will enable an SNR that is competitive with the auto-correlations, and jitter values down to 100 ps (which have been demonstrated (Mena-Parra *et al.*, 2022)) will permit SNR values below 0.1, which is an SNR region inaccessible to the auto-correlations that we expect to be relevant for most sidelobe measurement regimes. In addition, we have demonstrated that we can achieve good phase recovery even at large jitter values of ∼3 ns. This compels us to target clocks that have at most 1.7 ns jitter, and provides good evidence to develop a system with 100 ps jitter.

## 3. LimeSDR Implementation and Preliminary Results

We have demonstrated the feasibility of the deterministic digital calibrator source approach using a commercially available LimeSDR software-defined radio (SDR) board (LimeSDR, 2016) that uses open-source software, GNURadio (GNURadio, 2001). Although the LimeSDR boards do not have wide enough bandwidth to be used directly by any 21 cm cosmology instrument, they enable an efficient path for development and testing. As described below, first we verified the noise generation and correlation internally within a single LimeSDR board, and then we used two boards to form a beam map of a radio antenna to compare against a vector-network-analyzer (VNA) measurement.

The techniques presented here, while used with narrow bandwidths, can be applied in the same way to a wide bandwidth instrument by first using a coarse channelization, which for cosmology applications is approximately 500 kHz, and treating each complex frequency channel independently. This allows for precise spectral calibration across a large bandwidth using these same techniques.

### 3.1. *LimeSDR Benchtop validation*

We first assessed the signal-to-noise and phase retrieval capabilities of the LimeSDR using a single LimeSDR board, using it to produce a random sequence in one LimeSDR input and the same random sequence with additional Gaussian noise in a second input, as shown schematically in Fig. 4(a). Using GNURadio, we generated pseudo-random Gaussian noise, which is passed through a DAC and transceiver to be transmitted at a central frequency of 1420 MHz with a bandwidth of 7 MHz. The signal is attenuated and sent through a Low Noise amplifier, passed through another transceiver, and digitized. The resultant signal is then correlated with the original digital signal ("correlated") and with itself ("autocorrelation"). The analog filters in the transceiver chain of the LimeSDR board cut off within the measurement bandwidth shown here, so we expect good results from ∼1418 to 1422 MHz. The source is also turned off to allow an assessment of both signal and noise.

First, a small FFT is performed on the correlation spectrum, which produces a lag spectrum (Fig. 4(b)). The spectrum has a peak at ∼5 ms, which demonstrates both that we have measured a correlated signal, and also that there is a significant time delay between the two LimeSDR channels. This can be attributed to delays in software initialization of the LimeSDR board. The size of the spike at a single lag as compared to the rest is an estimate of the signal-to-noise in the cross correlation (the white noise is the uncorrelated noise level, the spike is the signal level at the lag). The signal is isolated to the peak in the lag spectrum, and the real and imaginary components can be extracted from the peak and used to measure the phase (Fig. 4(c)) and amplitude (Fig. 4(e)) of the correlated signal. As expected, the phase is flat across the bandpass, consistent with a single delay within the analog filtering. The amplitude of the correlated spectrum is nearly two orders of magnitude higher than the noise ("off") signal. The auto-spectrum (Fig. 4(d)) also shows the signal on and off data sets, where the excess power above noise is about 0.7. The equivalent quantity for the correlated data set is two orders of magnitude higher.

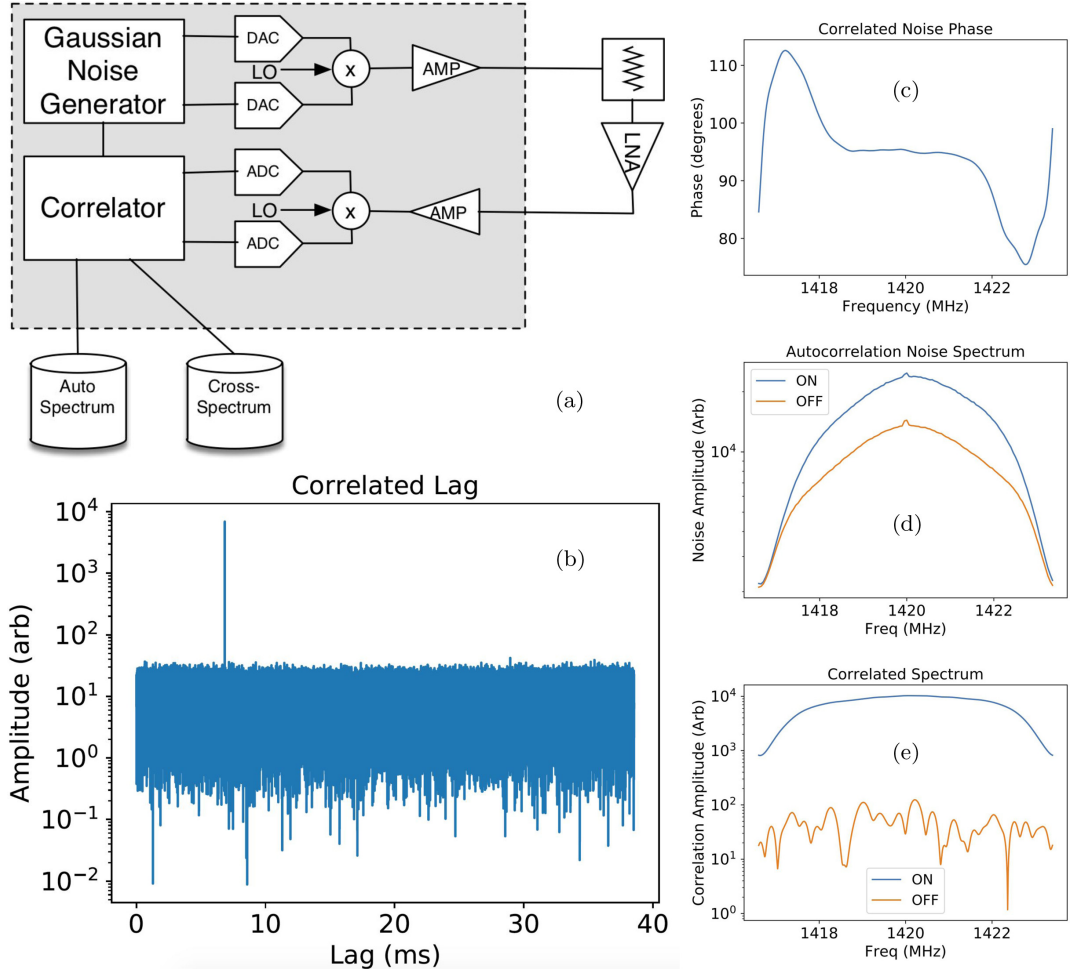

Fig. 4. (Color online) (a) Schematic of the pseudo random noise test using a single LimeSDR board and GNURadio, where the test is described in more detail in the text. (b) Correlated pseudo-random signal against the transmitted, attenuated and amplified signal. (c) Phase retrieval from the correlation of pseudo-random noise and the transmitted, attenuated and amplified signal. (d) Autocorrelation of pseudo-random noise transmitted, attenuated and amplified. Comparison of signal (blue), to the same processing with the transmission disconnected (orange). (e) The amplitude of the correlated signal, with the same colors as (d).

This test demonstrated that we can use a LimeSDR board and GNURadio software to measure a correlated signal in amplitude and phase, and with higher signal to noise than in the autocorrelation channel.

### 3.2. *Antenna beam map with a LimeSDR calibration source*

To test the feasibility of a free-space deterministic noise source, we used the testing setup shown in Fig. 5(a) to measure the angular response (beam pattern) of a feed and compare the results with a VNA measurement. This test was performed outside in an environment with significant RFI, with the software code modified to operate at ∼420 MHz. The signal was generated by a LimeSDR using the GNUradio Gaussian noise source, which generates a pseudo-random sequence using the xoroshiro128 uniform generator and passes it through a transformation to make it Gaussian. The signal is then broadcast from a wide-band cloverleaf antenna similar to those used for the CHIME experiment (Deng & Campbell-Wilson, 2014) with an additional cylindrical choke to symmetrize the beam. The signal was received by a second cloverleaf antenna and detected by a second LimeSDR system that is also generating an identical signal pattern.

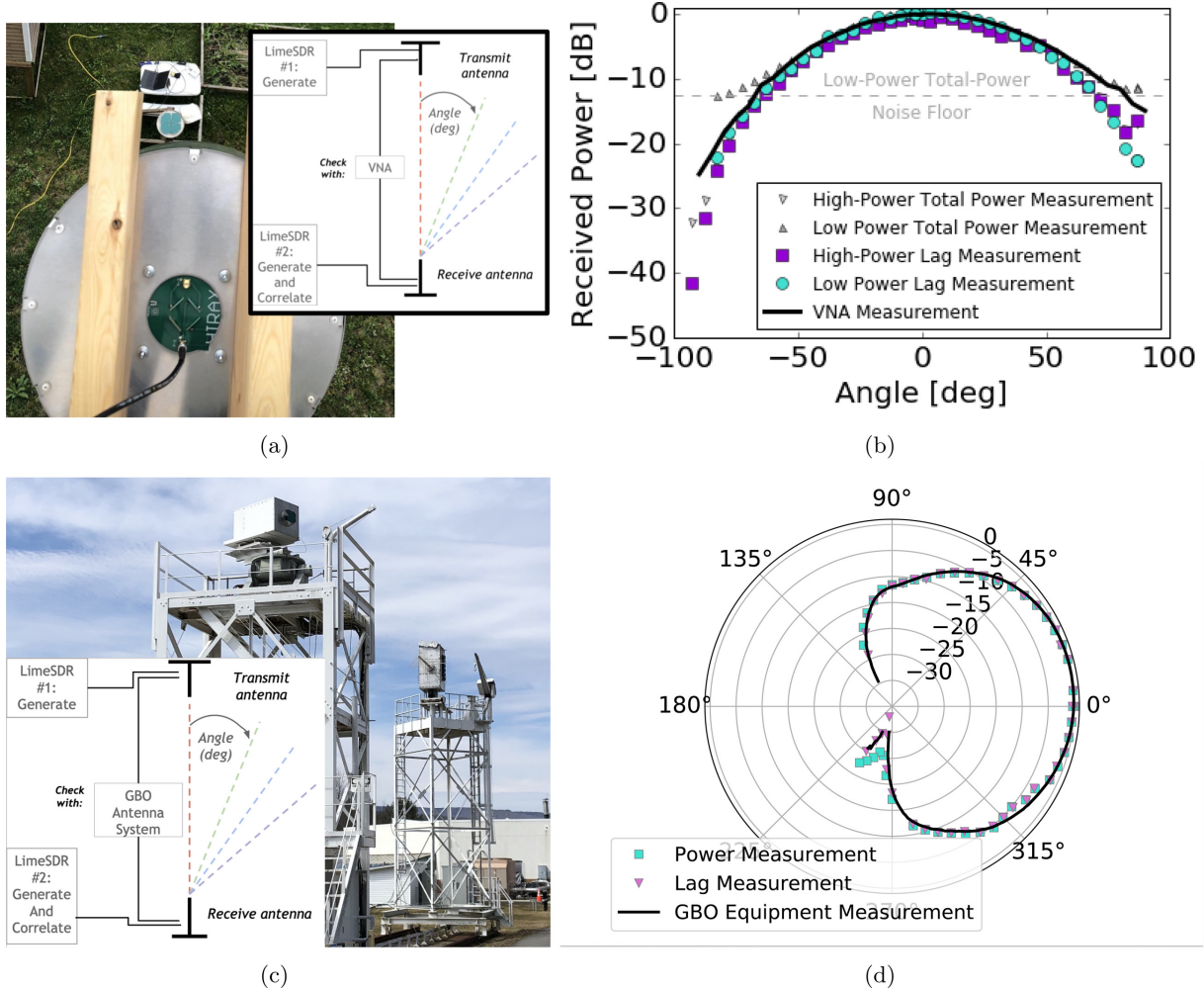

Fig. 5. (Color online) (a) Measurement photo looking from behind transmit antenna and schematic of setup at ∼420 MHz. (b) Preliminary comparison of antenna beam measurements between a VNA and the known noise source with high-power transmission and low-power transmission. The cyan and purple points measure the beam pattern locking into the max amplitude lag signal, measuring the cross-correlation. The gray points measure the total power received as a function of angle. The low-power total power shows the thermal noise in the autocorrelation contributing to a noise floor. (c) Measurement photo and schematic of setup at the GBO test range facility at ∼500 MHz. (d) Preliminary comparison of antenna beam measurements between the GBO antenna range measurement and the deterministic noise source. The purple points measure the beam pattern locking into the max amplitude lag signal, measuring the cross-correlation. The cyan points measure the total power received as a function of angle. The total power shows the thermal noise in the autocorrelation contributing to a noise floor, deviating from the GBO measurement in the backlobe where the signal is lowest.

The measured signal is detected in auto-correlation and cross-correlation with the second LimeSDR. Because the LimeSDR clock was known to fail the timing requirements, and two precision GPS-disciplined oscillators were not available for this test, the transmitter and receiver shared the same 10 MHz clock source, but generated the pseudo-random sequence independently.

The receiving antenna is rotated on its platform, and the auto- and cross-correlation spectra are recorded along with the angle of rotation to form a measurement of the receiving antenna's beam. The angle was measured using a digital inclinometer. We expect the beam to be broad, similar to a dipole beam. The beam mapping results are shown in Fig. 5(b) showing both auto-correlation (labeled "Total Power") and cross-correlation (labeled "Lag") as a function of antenna angle for three scenarios: broadcasting a high power signal (approximately −10 dBm before cable and antenna losses), a low power signal (approximately −40 dBm, and a VNA measurement (approximately 0 dBm). These results show the following features: in both cases (auto- and cross-correlation), a high powered signal allows an improved measurement at large angles where the beam response is low; in the case of a low-powered signal, the auto-correlation measurement is limited by the system noise while the cross-correlation measurement is not; and all are broadly consistent with the VNA measurement, although discrepancies arise at larger angles, ascribed to imperfect repeatability in the axis perpendicular to the measured angle (small changes in the elevation cut can have large sidelobe changes). This shows that the cross-correlation SNR was as good or better than the auto-correlation measurement and was never limited by a noise floor. Although the cross-correlation has the additional benefit of measuring phase, as discussed below, the LimeSDR pair were not sufficient to demonstrate a measurement of beam phase with adequate signal-to-noise. The results shown here demonstrate the first free-space correlate-able calibrator signal deterministically generated from time stamps.

### 3.3. *Antenna beam mapping results from GBO antenna range*

After demonstrating the benefit and feasibility of the free-space calibration source, we repeated the measurements at the outdoor test range at Green Bank Observatory (Fig. 5(c)). This allowed us to compare the LimeSDR source with a professional beam mapping facility in an RFI clean environment. We again modified the LimeSDR code to operate at 500 MHz to match the frequency of the antenna range. As before, a common clock was distributed to both LimeSDRs and a signal was transmitted from a stationary antenna, the signal is detected in a second rotate-able antenna, and the beam pattern of the receiver antenna was measured. The measurement was performed at 500 MHz with a common 10 MHz clock. The results are shown in Fig. 5(d), showing that the free-space deterministic LimeSDR source is consistent with the GBO measurements at 500 MHz. There is an improved agreement over the preliminary measurements at large angles ($> 50°$) and the back-lobe of the antenna ($> \pm 90°$) could be measured. As before, the SNR was far better than an auto-correlation measurement at lower powers (most apparent in the back lobe), indicating that the noise floor from the auto-correlation can impede the full beam measurement. However, there is some discrepancy around $320°$, due to a restart of the equipment, and insufficient time allowed for temperature stabilization.

The asymmetry of the pattern, in particular the excess power in one half of the backlobe region, is a known artifact due to the presence of buildings to the right of the test setup, and can be seen in the GBO test range equipment measurement as well as the coherent LimeSDR source measurements.

### 3.4. *LimeSDR calibration source limitations*

During the course of measurement we also found a significant and unstable ~5 ms lag between input channels, which is longer than the typical sampling time for most applications of the this calibration signal. The measurements presented above include a free-space demonstration, however we had to use a common clock distributed directly to the LimeSDR boards, instead of a pair of GPS-disciplined clocks. In addition, during the course of testing with the LimeSDR boards and GNURadio, we found that packets were silently dropped which resulted in reduced SNR. We also found suppressed performance near the center frequency and a strong gain drift with temperature. As a result, the coherent sources require significant but feasible improvements above the capabilities of the LimeSDR boards before they can be deployed for calibration purposes.

## 4. Summary and Outlook

In this paper, we have described a digital free-space coherent radio source that can be correlated in radio telescope correlators suitable for use as a calibration signal, and quantified the requirements on gain and timing. We have shown results from an initial implementation with commercial, low-bandwidth boards, and we have also provided the first demonstration that this signal can be transmitted and received in free space by measuring a radio antenna beam pattern at two frequency ranges (420 MHz on campus, and 500 MHz at the GBO). We also found that the LimeSDR implementation was insufficient for use as a calibration source, in particular it required us to distribute a common clock to the two calibration boards.

Precise clocking, high precision and high bandwidth DACs, and pseudo-random noise generation with minimal correlation at any reasonable lag are required to enable the calibration of radio telescopes for 21 cm cosmology with a digital noise source. Modern FPGAs and DACs have progressed to the point where the precision required is now possible, for example the Xilinx Zync Ultrascale+ RFSoC (RFSoC, 2017).

Finally, we note that the key generator could be implemented in the correlator FPGA itself by generating the digital sequence on the fly with a GPS timestamp such that the second board at the correlator would not be necessary. This provides a variety of new calibration opportunities, including the ability to run many different versions of the pseudo-random sequence at the same time. This would allow you to fly multiple sources for more efficient beam mapping. In future instruments, if the digitization occurs directly at the focus of the instrument, these sequence generators could be running on each separate antenna, allowing for independent gain calibration of each antenna separately, at any cadence.

## Acknowledgments

We are grateful for helpful discussions with Keith Vanderlinde. This material is based upon work supported by the National Science Foundation under Grant Nos. 1751763, 2107929, and 2108338.